\pdfoutput=1
\documentclass[twocolumn,prl,superscriptaddress,longbibliography]{revtex4-2}
\usepackage[colorlinks=true, citecolor=blue, urlcolor=blue, linkcolor=red]{hyperref}
\renewcommand{\section}[1]{{\par\it #1.---}\ignorespaces}
\usepackage{amsmath,amssymb,tikz,scalerel}
\hypersetup{
        colorlinks=true,
        linkcolor=red,
        citecolor=blue,
        urlcolor=blue}
\usetikzlibrary{svg.path}
\definecolor{orcidlogocol}{HTML}{A6CE39}
\tikzset{
	orcidlogo/.pic={
		\fill[orcidlogocol] svg{M256,128c0,70.7-57.3,128-128,128C57.3,256,0,198.7,0,128C0,57.3,57.3,0,128,0C198.7,0,256,57.3,256,128z};
		\fill[white] svg{M86.3,186.2H70.9V79.1h15.4v48.4V186.2z}
		svg{M108.9,79.1h41.6c39.6,0,57,28.3,57,53.6c0,27.5-21.5,53.6-56.8,53.6h-41.8V79.1z M124.3,172.4h24.5c34.9,0,42.9-26.5,42.9-39.7c0-21.5-13.7-39.7-43.7-39.7h-23.7V172.4z}
		svg{M88.7,56.8c0,5.5-4.5,10.1-10.1,10.1c-5.6,0-10.1-4.6-10.1-10.1c0-5.6,4.5-10.1,10.1-10.1C84.2,46.7,88.7,51.3,88.7,56.8z};}}
\newcommand\orcid[1]{\href{https://orcid.org/#1}{\mbox{\scalerel*{\begin{tikzpicture}[yscale=-1,transform shape]\pic{orcidlogo};\end{tikzpicture}}{|}}}}

\begin{document}
\title{Floquet topological phases with time-reversal and space inversion symmetries and
dynamical detection of topological charges}
\author{Hong Wu\orcid{0000-0003-3276-7823}}\email{wuh@cqupt.edu.cn}
\affiliation{School of Science, Chongqing University of Posts and Telecommunications, Chongqing 400065, China}
\author{Yu-Chen Dong}
\affiliation{School of Science, Chongqing University of Posts and Telecommunications, Chongqing 400065, China}
\author{Hui Liu}\email{hui.liu@fysik.su.se}
\affiliation{Department of Physics, Stockholm University, AlbaNova University Center, 10691 Stockholm, Sweden}

\begin{abstract}
For spinful systems with spin 1/2, it is generally believed that $\mathcal{P}$ and $\mathcal{T}$ invariant strong and
second-order topologies exist in four band and eight band system, respectively. Here, by using periodic driving, we find it is possible to have strong topological insulator, second-order topological insulator and hybrid-order topological insulator in a single four band system. Furthermore, we established a direct connection between topology and dynamics. More convenient experimental detection for these topological phases has also been proposed. This study provides the theoretical basis for novel topological insulator that possess hybrid-order boundary states beyond the conventional regimes.

\end{abstract}

\maketitle
\textcolor{blue}{\section{Introduction}}
Various topological phases have garnered significant interest, including topological insulators \cite{RevModPhys.83.1057,RevModPhys.88.021004,RevModPhys.82.3045}, Weyl/Dirac semimetals \cite{RevModPhys.90.015001,PhysRevX.5.031013,PhysRevLett.108.266802,PhysRevLett.119.076801,PhysRevLett.118.106406,PhysRevLett.129.246404,PhysRevLett.114.225301}, topological superconductivity \cite{PhysRevLett.126.137001,PhysRevLett.115.187001}. 
In recent years, the notion of n-dimensional (nD) topological phases with (n-m)D topological boundary states, known as m-order topological phases, have been proposed. For example, 3D higher order topological insulators with gapless hinge states \cite{Benalcazar_2017,Schindler_2018,PhysRevX.9.011012,PhysRevLett.118.076803,PhysRevLett.119.246402,PhysRevLett.126.206404} and higher order topological semimetals featuring hinge Fermi arcs \cite{PhysRevLett.125.266804,PhysRevLett.125.146401,PhysRevB.104.014203,PhysRevB.105.224312,PhysRevB.105.L081102,PhysRevB.105.L060101,Wei_2021,PhysRevLett.128.026405,PhysRevB.106.035105,PhysRevResearch.3.L032026,PhysRevLett.127.196801,PhysRevB.104.014203,PhysRevB.105.L081102,PhysRevB.107.085108,mao2023higherorder} have been predicted.
Both various theoretical schemes to realize these new phases \cite{PhysRevLett.120.026801,PhysRevB.104.224303,PhysRevB.101.245110,PhysRevB.103.115428,PhysRevB.100.245108,PhysRevLett.124.036401,PhysRevLett.123.177001,PhysRevLett.123.036802,PhysRevB.98.205129,PhysRevLett.128.127601,PhysRevB.101.235403,PhysRevB.104.165416,PhysRevResearch.2.023115,PhysRevLett.123.216803,PhysRevResearch.3.L042044,PhysRevB.97.205136,PhysRevB.104.L121108,PhysRevLett.126.016802,PhysRevLett.125.097001,PhysRevResearch.2.043131,PhysRevResearch.4.023049,PhysRevLett.123.247401,PhysRevA.101.043833,PhysRevB.105.045417,PhysRevB.99.020508,PhysRevB.102.165153,PhysRevResearch.2.013348} and some experimental observations \cite{PhysRevLett.125.213901,PhysRevLett.126.146802,PhysRevLett.122.204301,PhysRevLett.125.255502,PhysRevLett.122.233903} are reported over the past few years. 

All topological phases have been classified by considering various symmetries. Due to the fundamental roles
in solid quantum materials, photonic systems, and classical acoustic systems, the physics of time-reversal $\mathcal{T}$ and space inversion $\mathcal{P}$ symmetry have attracted much attention. For spinful systems with spin $1/2$, $\mathcal{P}$ and $\mathcal{T}$ can lead to a Kramers double degeneracy at every momentum. The representative topological phases are Dirac semimetals and topological insulator with helical boundary states \cite{RevModPhys.83.1057}. For spinless system,  $\mathcal{P}$ and $\mathcal{T}$ guarantees a real band structure. The unique topological phases for this class are Stiefel-Whitney semimetal and insulator \cite{PhysRevLett.132.197202,PhysRevLett.121.106403}. These states show possibilities in robust high-Q-resonance-based sensing, energy harvesting, and spin electric devices \cite{RevModPhys.83.1057,PhysRevLett.132.197202}. Here, We focus on spinful systems. In four band system with the $\mathcal{P}$ and $\mathcal{T}$ invariant
first order topological insulator, there are gapless helical
edge/surface states in the gap. By breaking $\mathcal{T}$ symmetries, the one dimensional gapless helical boundary modes will be gapped out in a nontrivial way and the first order topological phase is transited to a second order topological phase \cite{PhysRevB.100.115403}. The combination of two such modified models can generate a eight band model for $\mathcal{P}$ and $\mathcal{T}$ invariant second order topological insulator \cite{Schindler_2018}. It seems that there is a fundamental limitation that does not allow first order and second order topological insulators to coexist in a four band system. This impedes the development of $\mathcal{P}$ and $\mathcal{T}$ class-related topological physics and corresponding applications. It’s natural for us to wonder whether the $\mathcal{P}$ and $\mathcal{T}$ symmetry induced conventional limitation is always true in various systems.

Coherent control via periodic driving dubbed Floquet engineering has become a versatile tool in artificially creating novel topological phases in systems of ultracold atoms \cite{RevModPhys.89.011004,PhysRevLett.116.205301}, photonics \cite{Rechtsman2013,PhysRevLett.122.173901}, superconductor qubits \cite{Roushan2017}, and graphene \cite{McIver2020}. Inspired by that periodic driving can generate new topological phases without static analogs in a controllable manner \cite{PhysRevLett.124.057001,PhysRevB.103.L041402,PhysRevB.100.085138,PhysRevResearch.1.032045,PhysRevB.107.235132,PhysRevA.100.023622,PhysRevB.106.184106,PhysRevB.109.035418}, a natural question is whether the new $\mathcal{P}$ and $\mathcal{T}$ invariant topological phases without conventional limitations can be induced by periodic driving. If yes, how to exactly characterize and detect these new states?
  
\begin{figure}[tbp]
\centering
\includegraphics[width=1\columnwidth]{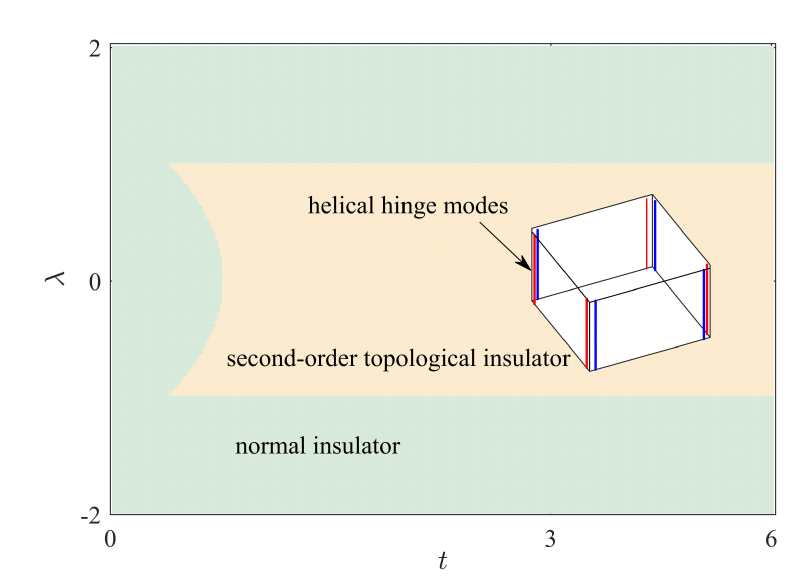}
 \caption{ Phase diagram in $\lambda-t$ plane. The inset is a schematic illustration of the distribution of four branches of helical hinge modes. 
}
\label{shiyi}
\end{figure}

Motivated by these considerations, we propose a scheme to artificially create $\mathcal{P}$ and $\mathcal{T}$ invariant topological phases by Floquet engineering. The strong topological insulator, second-order topological insulator and hybrid-order topological insulator can be studied in a single four band system.  The corresponding complete description and understanding of these topological phases have also been established. A direct connection between topological invariants and dynamics has also been discovered. Based on this,  more convenient experimental detection of these topological phases have also been proposed. This study provides a theoretical basis for subsequent experimental research. All results open an unprecedented possibility to realize new phases without static analogs.

\textcolor{blue}{\section{Static topological phases}}
We consider a 3D second-order topological insulator whose Bloch Hamiltonian reads
\begin{equation}
\mathcal{H}_1(k_x,k_y,k_z)=f_1\sigma_x\tau_x+f_2\sigma_x\tau_z+f_3\sigma_y\tau_0+f_4\sigma_z\tau_0.
\end{equation}
where $\sigma_x$ and $\tau_i$ are Pauli matrices acting on spin and orbital degree of freedoms, respectively. This system respects time-reversal, space inversion, mirror chiral symmetry, and chiral symmetries.
\begin{eqnarray}
&\mathcal{P} \mathcal{H}_1(\mathbf{k}) \mathcal{P}^{-1}=\mathcal{H}_1(-\mathbf{k}),\,\,\,\,\, \mathcal{T} \mathcal{H}_1(\mathbf{k}) \mathcal{T}^{-1}=\mathcal{H}_1(-\mathbf{k}) \label{dui1}\\&\mathcal{M}\mathcal{H}_1(k_x,k_y,k_z)\mathcal{M}^{-1}=-\mathcal{H}_1(k_y,k_x,k_z) , \label{dui2} \\&\mathcal{C}\mathcal{H}_1(k_x,k_y,k_z)\mathcal{C}^{-1}=-\mathcal{H}_1(k_x,k_y,k_z) , \label{dui3}
\end{eqnarray}
where $\mathcal{T}=i\sigma_0\tau_y\mathcal{K}$ and $\mathcal{P}=\sigma_z\tau_0$ are the time-reversal and spatial-inversion operations, respectively. $\mathcal{M}=\sigma_x\tau_z$ and $\mathcal{C}=\sigma_x\tau_y$. Due to the mirror chiral symmetry and chiral symmetry which
ensures that the eigenvalues appear in ($E,E,-E,-E$) pairs and ($E(k_x,k_y)+E(k_y,k_x)=0$), Hamiltonian on the high-symmetry line $k_x=-k_y=k$ can be expressed as Diag[$\mathcal{H}_{1,+},\mathcal{H}_{1,-}$] with $\mathcal{H}_{1,\pm}={\bf h}_\pm\cdot{\pmb\sigma}=\pm f_2\sigma_x+f_3\sigma_y+f_4\sigma_z$. Then we can construct mirror spin Chern number $C_{xy}$, which can be used to characterize the topology of our system. The expression for $C_{xy}$ is given by $C_{xy}=\frac{C_{+}-C_{-}}{2}$ with
\begin{equation}
{C}_{\pm}=\frac{1}{4\pi}\int_\text{BZ}\frac{1}{E_{\pm}^{3/2}}\mathbf{h}_{\pm}\cdot(\partial_{k}\mathbf{h}_{\pm}\times\partial_{k_z}\mathbf{h}_{\pm})d{k}d{k_z},
\end{equation}
where $E_{\pm}$ are the eigen energies of $\mathcal{H}_{1,\pm}$. If the change of $C_{xy}$ is 2, there is a phase transition from trivial insulator to second-order topological insulator. Although mirror chiral symmetry and chiral symmetry are used to construct our unique topological insulator. it is itself not essential for the existence of topological boundary states . 

Without loss of generality, we choose $f_1=\sum_i 2(\cos k_i+\lambda)\sin k_i$, $f_2=2[(\cos k_x+\lambda)\sin k_y-(\cos k_y+\lambda)\sin k_x  ]$, $f_3=\sin k_z$, and $f_4=\sum_i[(\cos k_i+\lambda)^2-\sin^2k_i]-t(\cos k_z-1)$. Here $i=x,y$. Note that the counterpart with the same symmetries was investigated in Ref. \cite{PhysRevLett.123.177001}. When $t>t_c=1-\frac{\lambda^2}{2}$ and $\lvert \lambda \lvert<1$, $C_{xy}=2$ signifies the formation of a second-order topological insulator with antipropagating Kramers pairs of hinge modes (see Fig. \ref{shiyi}). The underlying mechanism for the formation of hinge states can be explained by edge theory. As for the 2D subsystem $\mathcal{H}_1(k_z=0)$, Ref. \cite{PhysRevLett.123.177001} has demonstrated that a generic edge of this 2D system is a 1D insulator described by the gapped Dirac model. It is well known that this kind of edge model has a $Z_2$ topological classification by the sign of the mass term.  When the mass terms of two neighboring edges have opposite signs, two neighboring edges belong to distinct $Z_2$ classes and Jackiw-Rebbi theory suggests that protected 0D corner mode must exist at the intersection between two edges \cite{PhysRevB.104.085205}. When we make a Taylor expansion at the $k_z=0$, the surface model of $\mathcal{H}_1$ can be written as $\mathcal{H}_{\text{surface},3D}=\mathcal{H}_{\text{edge},2D}+k_z\Gamma$ where $\mathcal{H}_{\text{edge},2D}$ is the edge model of 2D subsystem $\mathcal{H}_1(k_z=0)$ with $\{\mathcal{H}_{\text{edge},2D},k_z\Gamma\}=0$. Obviously, the mass terms of adjacent surfaces have opposite sign. The hinge is a domain wall of Dirac mass and consequently harbors hinge state.

\textcolor{blue}{\section{Floquet topological phases}}We consider a periodically driven system whose Bloch Hamiltonian is
\begin{equation}
\mathcal{H}({\bf k},t)=\mathcal{H}_1({\bf k})+\mathcal{H}_2({\bf k})\delta(t/T-n),\label{odr}
\end{equation}
where $\mathcal{H}_2=m_z\sigma_z\tau_0$, $T$ is the driving period, and $n$ is an integer. It's easy to find that both $\mathcal{H}_1$ and $\mathcal{H}_2$ posses $\mathcal{P}$ and $\mathcal{T}$ symmetry. Due to the fact that energy of our system is not conserved, this kind of time-periodic system does not has well-defined energy spectrum.   According to Floquet theorem, the one-period evolution operator ${U}(T)=\mathbb{T}e^{-i\int_{0}^{T}\mathcal{H}(t)dt}$ defines an effective Hamiltonian $\mathcal{H}_\text{eff}\equiv {i\over T}\ln {U}(T)$ whose eigenvalues are called the quasienergies \cite{PhysRevA.91.052122}. From the eigenvalue equation $U({T})\lvert u_l \rangle=e^{-i\varepsilon_{l}T}\lvert u_l \rangle$, we conclude that quasienergy $\varepsilon_l$ is a phase factor, which is defined modulus $2\pi/T$ and takes values in the first quasienergy Brillouin zone [$-\pi/T$,$\pi/T$] \cite{PhysRevLett.113.236803}. The topological phases of our periodically driven system are defined in such quasienergy spectrum \cite{PhysRevB.108.054310}. Although the symmetries of $\mathcal{H}_\text{eff}\equiv {i\over T}\ln {U}(T)$ may be different from those of the original static system, all the symmetries can be inherited by $\mathcal{H}_{eff}$ in symmetric time frame. This frame is obtained by shifting the starting time of the evolution backward over half of the driving period. The resulting new effective Hamiltonian is $\mathcal{H}'_{eff}=\frac{i}{T}\ln[e^{-i\mathcal{H}_1T/2}e^{-i\mathcal{H}_2T}e^{-i\mathcal{H}_1T/2}]$. The general form of effective Hamiltonian is
$
\mathcal{H}'_{eff}=f'_1\Gamma_1+f'_2\Gamma_2+f'_3\Gamma_3+f'_4\Gamma_4$, where $\Gamma_1=\sigma_x\tau_x$, $\Gamma_2=\sigma_x\tau_z$, $\Gamma_3=\sigma_y\tau_0$ and $\Gamma_4=\sigma_z\tau_0$ (see proof in the Supplemental Materials). According to Supplemental Material, band touching point occurs when the parameters satisfy either
\begin{eqnarray}
&2(1-\lambda^2)+t(e^{i\alpha_z}-1)-m_z=n_{\alpha_z}\pi/T& \label{btps1}\\
&\text{or}~~2[1+\lambda^2+\lambda(e^{i\alpha_x}+e^{i\alpha_y})]-t(e^{i\alpha_z}-1)\nonumber\\&+m_z=n_{\alpha_x,\alpha_y,\alpha_z}\pi/T&, \label{btps2}
\end{eqnarray}
at the quasienergy zero ($\pi/T$) for even (odd) $n_{\alpha_z}$ and $n_{\alpha_x,\alpha_y,\alpha_z}$, with $\alpha_{x}=k_x$, $\alpha_{y}=k_y$, and $\alpha_{z}=k_z$ are $0$ or $\pi$. EQ. \eqref{btps1} and \eqref{btps2} supply a guideline to manipulate topological phase transition.

\begin{figure*}
\centering
\includegraphics[width=2.\columnwidth]{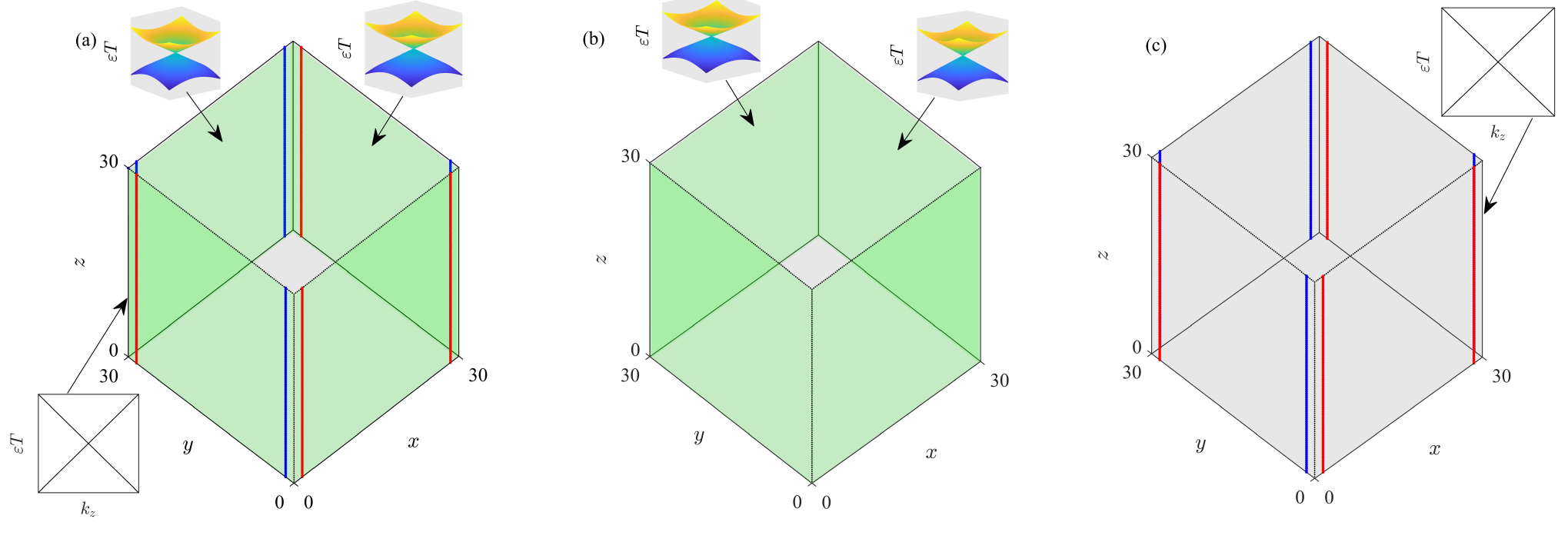}
\caption{ Schematic of the boundary states at quasienergy zero and $\pi/T$ gap. (a) The system harbors antipropagating Kramers pairs of hinge modes (red/blue lines) and gapless surface Dirac cone (colored in green) at quasienergy zero and $\pi/T$ gap, respectively. (b) The system only hosts  gapless surface Dirac cone (colored in green) at $\pi/T$ gap. (c) The system only hosts antipropagating Kramers pairs of hinge modes (red/blue lines) at zero gap. We use $m_z=1$, $T=0.3$ in (a), $m_z=\frac{5}{3}$, $T=0.3$ in (b) and $m_z=0.3$, $T=0.1$ in (c).  The inset of (a), (b), and (c) show the energy spectrum of hinge states or surface Dirac states.
 The other parameters are $\lambda=-0.5$ and $t=4$.
} \label{energy}
\end{figure*}

Different from previous results in static systems, $\mathcal{P}$ and $\mathcal{T}$ invariant strong topological insulator , hybrid topological insulator can also be studied in our Floquet system. To give a complete description for the Floquet system, $Z_4$ index can be defined to characterize first order strong topological insulators. 
\begin{equation}
Z_4=\frac{1}{4}\sum_{K\in \text{TRIMs}}({n^{+}_{K}-n^{-}_{K}})\,\,\,\,\,\,\,\text{ mod}\,\, 4,
\end{equation}
Where $n^{\pm}_{K}$ is the number of occupied states with even/odd parity at time reversal invariant momenta (TRIM). $Z_4=1$ or $3$ indicates a strong topological insulator \cite{PhysRevB.101.245110}.

Fig. \ref{energy} shows the topological phases in different $m_z$ and $T$.  Fig. \ref{energy}(a) shows the  schematic of gapless helical hinge states and surface Dirac cone in $0$ and $\pi/T$ gap, respectively (see Supplemental Material for quasienergy spectrum). The existence of first-order strong topological insulator can be predicted by $Z_4=1$. Then there are an odd number of Dirac cones on all other surfaces of the crystal. The contribution of strong topological insulator to Chern number $C_{xy}$ is 1. Due to the coexistence of first-order strong and second-order topological insulators, we obtain $C_{xy}=3$. By combining these two topological invariants, we can establish a complete description to the topological
phases in the periodically driven $\mathcal{P}$ and $\mathcal{T}$ invariant system. Here, such a composite topological insulator is fundamentally different from traditional $\mathcal{P}$ and $\mathcal{T}$ invariant topological insulators. Due to the fact that a general approach to realize second order topological phases in $\mathcal{P}$ and $\mathcal{T}$ invariant system is "order transition", that is, by breaking time reversal symmetry , the one-dimensional lower boundary modes will be gapped out in a nontrivial way, and accordingly, the first-order topological phase is transited to a higher-order topological phase \cite{PhysRevB.97.205134,PhysRevLett.121.196801}, 
the coexistence of first-order and second-order
topological phases has not yet been observed in three dimensional four band $\mathcal{P}$ and $\mathcal{T}$ invariant system. Besides, single topological phase can also be observed in our system. Fig. \ref{energy}(b) shows surface Dirac cone in $\pi/T$ gap. $Z_4=1$ and $C_{xy}=1$ confirm this strong topological insulators in our system. The second-order Floquet topological insulator with helical hinge states only in $0$ gap is also observed in Fig. \ref{energy}(c). Similar to the static case, topological invariants for this case are $Z_4=0$ and $C_{xy}=2$. Such rich phenomenas provide a new platform for the study of topological states of matter.

\textcolor{blue}{\section{Dynamical characterization}}
Topological phase transition is accompanied by the closing and reopening of the bulk-band gap \cite{PhysRevB.104.205117}. This implies that the exploration of topology for these band touching points is sufficient. The topological charge of a band touching point is defined as the  Berry curvature flux threading a closed surface that encloses the band touching point:  $\mathcal{N}_{\pm}=\frac{1}{2\pi}\oint_S\Omega_{\pm}(\mathbf{k})\cdot \mathbf{dS}$, where $\Omega_{\pm}(\mathbf{k})=\nabla_{\mathbf{k}}\times \bf{A}_{\pm}(\mathbf{k})$ with $ \bf{A}_{\pm}(\mathbf{k})=-i\langle u_{\pm}\lvert\partial_{\mathbf{k}}\lvert u_{\pm}\rangle$  is the Berry curvature and $\mathbf{S}$ encloses band touching point \cite{PhysRevLett.108.266802,PhysRevLett.127.187002}. $\lvert u_{\pm}(\mathbf{k}) \rangle$ is the eigenstate of $\mathcal{H}'_{eff,\pm}$ with $\mathcal{H}'_{eff}(k,-k,k_z)=\text{Diag}[\mathcal{H}'_{eff,+},\mathcal{H}'_{eff,-}]$. The effective Hamiltonian on the high-symmetry line $k_x=-k_y=k$ can be expressed as $\mathcal{H}'_{eff,\pm}= \pm f'_2\sigma_x+f'_3\sigma_y+f'_4\sigma_z$ (see proof in the Supplemental Materials). Now the Stokes theorem can be applied for the integration over the closed surface,
we can obtain $ \mathcal{N}_{\pm}=\frac{1}{2\pi}\oint_{l} \bf{A}_{\pm}(\mathbf{k})\mathbf{dl}$ with $\mathbf{l}$ is the closed path that encloses the band touching point. For convenience, we choose a tiny circle parametrized by $\theta$ on plane $k_z=0$ or $k_z=\pi$.
Due to the time-reversal symmetry, $\mathcal{N}_{+}+\mathcal{N}_{-}=0$.
 Then we only focus on $\mathcal{N}_{+}=\int_{0}^{2\pi} \frac{f'_2\partial_{\theta}f'_4-f'_4\partial_{\theta}f'_2}{f'^2_2+f'^2_4}d\theta$=$\frac{1}{2\pi}\int_{0}^{2\pi} \frac{d\phi}{d\theta}d\theta$ with $\tan\phi=f'_2/f'_4$.  Then we will show that $\phi$ is unobservable, but its double angle can be visible.

we propose a new method based on quantum quenches to characterize the topological charge. 
The initial state is fully polarized along the $\Gamma_0$ axis and $\rho_0$ is the density matrix of the initial state. Then, the stroboscopic time-averaged spin textures read $\overline{\langle \Gamma_i(\mathbf{k}) \rangle}=\lim\limits_{N\rightarrow \infty}\frac{1}{N}\sum_{n=0}^{N}\text{Tr}[\rho_0(\mathbf{k})U^{\dag}(\mathbf{k},nT)\Gamma_iU(\mathbf{k},nT)]=\frac{-f'_{i}f'_{0}}{\varepsilon^2(\mathbf{k})}$. 
We choose $\Gamma_0=\Gamma_4$, then
\begin{equation}
\tan 2\phi=\frac{2\overline{\langle \Gamma_2 \rangle}}{1+2\overline{\langle \Gamma_4 \rangle}},
\end{equation}
where $\sin 2\phi=-2\overline{\langle \Gamma_2 \rangle}$ and $\cos 2\phi=-(1+2\overline{\langle \Gamma_4 \rangle})$. This suggests that $\mathcal{N}_{+}$ is the phase change of $\frac{1}{4\pi}$arg[$-2\overline{\langle \Gamma_2 \rangle}$$-i-2i\overline{\langle \Gamma_4 \rangle}]$  as $\theta$ goes from $0$ to $2\pi$. Compared with the results in Ref. \cite{PhysRevLett.125.183001}, Our scheme has established the relationship between topological charge and stroboscopic time-averaged spin textures without introducing the concept of band inversion surfaces.  These results provide us with a way to directly measure topological invariants in Floquet system. 

\begin{figure}[tbp]
\centering
\includegraphics[width=1.02\columnwidth]{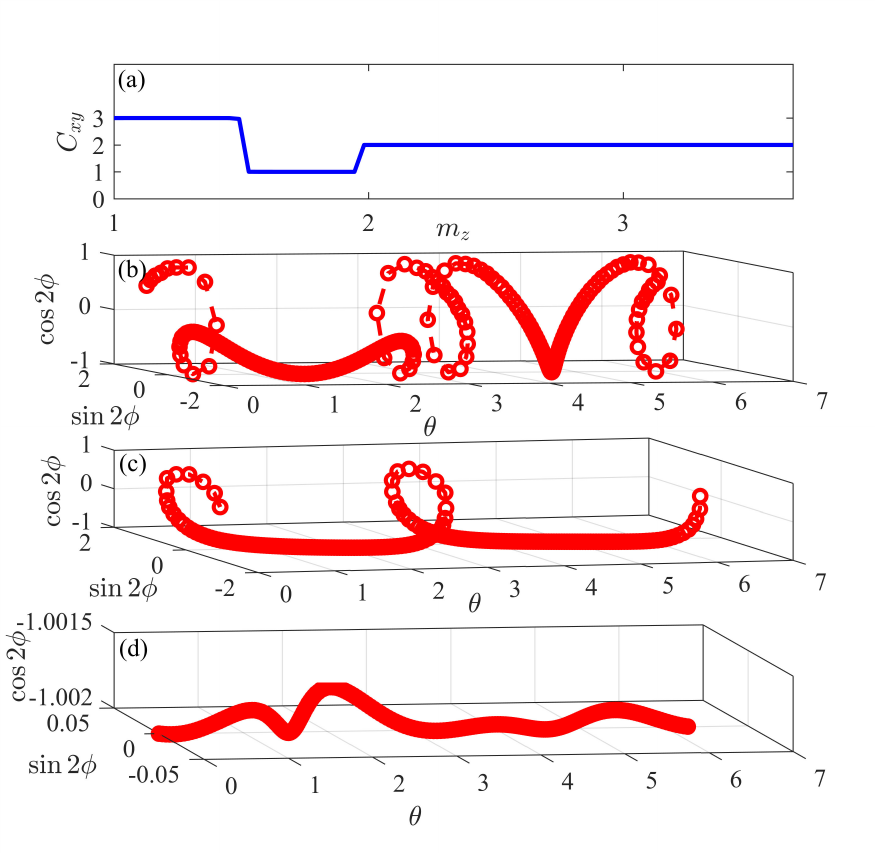}
 \caption{(a) Chern number with the change of $m_z$. (b)-(d) Trajectories of ($\theta$, $\sin 2\phi$, $\cos 2\phi$) with $\theta$ in $\mathbf{k}$ running from 0 to $2\pi$. $k_x=-k_y=k=1.1\cos\theta$, $k_z=0$ and $m_z=1.5+0.1\sin \theta$ in (a), $k_x=-k_y=k=0.1\cos\theta$, $k_z=\pi$ and $m_z=1.972+0.1\sin \theta$ in (c), $k_x=-k_y=k=0.1\cos\theta$, $k_z=\pi$ and $m_z=1+0.1\sin \theta$ in (d).
 We use $\lambda=-0.5$, $t=4$ and $T=0.3$.
}
\label{tance}
\end{figure}

 The Chern number at different $m_z$ is shown in Fig. \ref{tance}(a). According to EQ. \eqref{btps1} and \eqref{btps2}, the two band touching points in Fig. \ref{tance}(a) are located at $A'$: $k_x=-k_y=k=1.0472$, $k_z=0$ and $m_z=1.5$, $B'$: $k_x=-k_y=k=0$, $k_z=\pi$ and $m_z=1.972$, respectively. The respective changes in Chen numbers at these two band touching points are $-2$ and $1$, respectively. Then we plot the trajectories ($\theta$, $\sin 2\phi$, $\cos 2\phi$), calculated from stroboscopic time-averaged spin textures, in Fig. \ref{tance}(b), \ref{tance}(c) and \ref{tance}(d) . Fig. \ref{tance}(b) and (c) show that the change of Chern number at $m_z=1.5$ ($m_z=1.972$) is $\mathcal{N}_{+}=-2$ ($\mathcal{N}_{+}=1$). Besides, by choosing a trivial path on $m_z=1$ plane that encloses no band touching point, trajectories of ($\theta$, $\sin 2\phi$, $\cos 2\phi$) in Fig \ref{tance}(d) shows $\mathcal{N}_{+}=0$. This means that there is no change in the topological number at this point. The features in this quench dynamics match well
with the Chern number. Our results provide a basis for direct measurements and intuitive understanding of Floquet topological phase.

\begin{figure}[tbp]
\centering
\includegraphics[width=1.02\columnwidth]{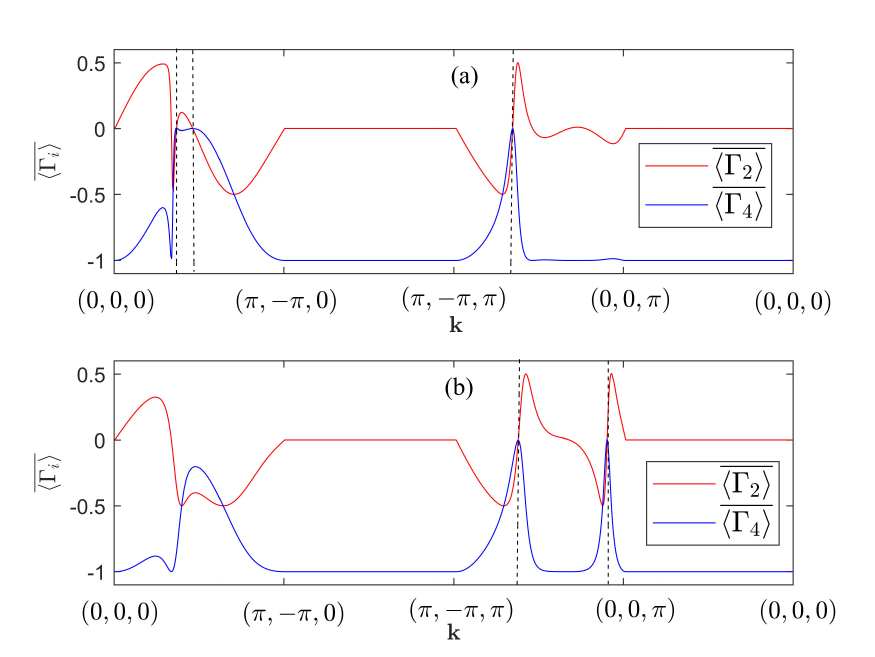}
 \caption{The stroboscopic time-averaged spin textures for (a) $m_z=1.64$ and (b) $m_z=2.31$.
 We use $\lambda=-0.5$, $t=4$ and $T=0.3$.
}
\label{tance2}
\end{figure}

Then, the $Z_4$ index, which is employed to identify strong topological insulator, can also be detected. 
For simplicity, we use $Z_2$ topological invariant to characterize strong topological insulator. It's given by  
\begin{equation}
(-1)^{Z_2}=\Pi_{\mathbf{k}_i}\text{sign}[{f'_4(\mathbf{k}_i)}],
\end{equation}
where $\mathbf{k}_i=$ is time reversal invariant momenta \cite{PhysRevLett.98.106803,PhysRevB.79.195322}. Due to the   mirror chiral symmetry $\sigma_x\tau_z\mathcal{H}'_{eff}(k_x,k_y,k_z)\sigma_x\tau_z=-\mathcal{H}'_{eff}(k_y,k_x,k_z)$, then we can obtain $(-1)^{Z_2}=\Pi_{\mathbf{k'}_i}\text{sign}[{f'_4(\mathbf{k'}_i)}]$ with $\mathbf{k'}_i\in (0,0,0)$, $(\pi,-\pi,0)$, $(\pi,-\pi,\pi)$, and $(0,0,\pi)$. To measure $Z_2$ index, we plot stroboscopic time-averaged spin textures in Fig. \ref{tance2}. When the red and blue lines intersect at zero, the sign of $f'_4$ will change. When $m_z=1.64$, Fig. \ref{tance2}(a) shows that $f'_4\lvert_{\mathbf{k}=(0,0,0)}$ ($f'_4\lvert_{\mathbf{k}=(\pi,-\pi,\pi)}$) and $f'_4\lvert_{\mathbf{k}=(\pi,-\pi,0)}$ ($f'_4\lvert_{\mathbf{k}=(0,0,\pi)}$) have same signs (opposite signs). The corresponding $Z_2=1$, reveals a strong topological insulator. Similarly, Fig. \ref{tance2}(b) reveals a Floquet topological phase with $Z_2=0$ at $m_z=2.31$. This phase is trivial in the first-order but nontrivial in the second-order ($C_{xy}=2$ is observed from Fig. \ref{tance}(a)). Here, the universal laws can also be observed. When the number of crossings between red line and blue line at zero is odd (even), $Z_2=1$ ($Z_2=0$). Up to now, we have established a direct connection between all topological invariants and dynamics.

\textcolor{blue}{\section{Discussion and Conclusions}}
Here, the delta-function driving protocol is considered just for the convenience of analytical calculation. Our scheme is generalizable to other driving forms, such as harmonic driving (see the detail in the Supplemental Materials). The three-dimensional higher-order topological insulator have been realized in phononic crystal \cite{PhysRevLett.127.255501} and photonic crystal \cite{PhysRevB.105.L060101,wang2024realization}. On the other hand, periodic driving has exhibited its power in engineering exotic phases in various experimental platforms such as, ultracold atoms \cite{RevModPhys.89.011004,PhysRevLett.116.205301}, superconductor qubits \cite{Roushan2017}, and photonics \cite{Rechtsman2013,PhysRevLett.122.173901}. By combining these developments, we believe that our proposal is achievable in the experiment.

We have investigated the topological insulators in $\mathcal{P}$ and $\mathcal{T}$ invariant systems. The  strong topological insulator, second-order topological insulator, and hybrid-order topological insulator can be studied in our four band system.  Such a phenomenon has not yet been observed. Then we proposed  a way to directly measure the topological invariants in Floquet system. Our results open an unprecedented possibility to realize new phases without static analogs.

\bibliography{references}
\end{document}